\documentclass[12pt]{article}
\usepackage{jheppub}
\usepackage{comment}
\usepackage{amsmath}
\usepackage{slashed}
\def\one{{\,\hbox{1\kern-.8mm l}}}
\newcommand{\Dslash}{\not{\hbox{\kern-4pt $D$}}}
\newcommand{\pdslash}{\not{\hbox{\kern-2pt $\partial$}}}

\newcommand{\Tr}{\mathrm{Tr}}

\newcommand{\La}{\Lambda}
\newcommand{\R}{\rho}

\newcommand{\Comment}[1]{{}}

\def\IZ{{\mathbb Z}}

\def\IR{{\mathbb R}}

\def\calh         {{\cal H}}

\def\calo         {{\cal O}}

\newcommand{\bc}{\begin{center}}
	\newcommand{\ec}{\end{center}}
\newcommand{\ba}{\begin{array}}
	\newcommand{\ea}{\end{array}}
\newcommand{\beq}{\begin{equation}}
	\newcommand{\eeq}{\end{equation}}
\newcommand{\bea}{\begin{eqnarray}}
	\newcommand{\eea}{\end{eqnarray}}
\newcommand{\bmx}{\begin{pmatrix}}
	\newcommand{\emx}{\end{pmatrix}}

\newcommand{\bt}{\beta}
\newcommand{\be}{\begin{equation}}
	\newcommand{\ee}{\end{equation}}
\newcommand{\dl}{\delta}

\newcommand{\ep}{\epsilon}

\newcommand{\la}{\lambda}
\newcommand{\T}{\tau}

\newcommand{\p}{\phi}
\newcommand{\s}{\sigma}

\newcommand{\Om}{\Omega}

\newcommand{\del}{\partial}
\newcommand{\half}{{\frac{1}{2}\,}}

\newcommand{\eref}[1]{Eq.\,(\ref{#1})}

\newcommand{\Ph}{\Phi}

\def\IB{\relax{\rm I\kern-.18em B}}
\def\IC{{\relax\hbox{\kern.3em{\cmss I}$\kern-.4em{\rm C}$}}}
\def\ID{\relax{\rm I\kern-.18em D}}
\def\IE{\relax{\rm I\kern-.18em E}}
\def\IF{\relax{\rm I\kern-.18em F}}
\def\II{\relax{\rm I\kern-.18em I}}
\def\IZ{\relax{\sf Z\kern-.35em Z}}
\def\Id{\relax{1\kern-.32em 1}}
\def\IG{\relax\hbox{$\inbar\kern-.3em{\rm G}$}}
\def\IR{\relax{\rm I\kern-.18em R}}

\newcommand{\ket}[1]{|#1 \rangle}
\newcommand{\bra}[1]{\langle #1 |}
\newcommand{\pa}{\partial}
\newcommand{\mH}{{\mathbb H}}
\newcommand{\mR}{{\mathbb R}}

\title{On phase transitions in the R\'enyi entropies of $2+1$-d large-N interacting vector models}

\author[a]{Harsha R. Hampapura,}
\author[a]{Albion Lawrence,}
\author[b]{Stefan Stanojevic}

\affiliation[a]{Martin A.Fisher School of Physics,\\
	Brandeis University, Waltham MA 02453}
\affiliation[b]{Department of Physics, Brown University, Providence RI 02912}

\emailAdd{hrharsha@brandeis.edu}
\emailAdd{albion@brandeis.edu}
\emailAdd{stefan\_stanojevic@brown.edu}

\abstract{We consider the reduced density matrix for a disc in the ground state of the interacting fixed points of the large-N O(N) vector model in $2+1$ dimensions. Using the map to the free energy on $\mH_2\times S^1$, we show that there is an instability in the R\'enyi entropies at $n = 1$, in the $N\to\infty$ limit, indicating a phase transition at or near this R\'enyi parameter. We close with a discussion of the finite-N case.}

\preprint{BRX-TH-6638}
\begin{document}
	
\maketitle	
\flushbottom
\section{Introduction}

Consider a quantum system with Hilbert space $\calh$, with a factorization $\calh = \calh_A \otimes \calh_B$. Given a pure state $\ket{\Psi}$ such as the ground state, the spectrum of the reduced density matrix $\rho_A = \Tr_B \ket{\Psi}$ is a measure of the entanglement between $\calh_A$ and $\calh_B$ in the state $\psi$. This can contain much useful information about the system at hand, particularly for quantum field theories. A judicious choice of factorization provides order parameters sensitive to zero-temperature quantum phase transitions \cite{hamma2005ground,Kitaev:2005dm,Levin:2006zz,Flammia:2009axf}. In the case of 2d conformal field theories, there is some evidence that the entanglement spectrum corresponding to the ground state, for a suitable set of factorizations, can completely characterize the theories \cite{Headrick:2012fk}, at least away from limits with duals described by classical Einstein gravity.  In the case of theories with a dual gravitational description, the entanglement properties of a given state appear to be key to constructing the dual geometry \cite{VanRaamsdonk:2010pw,Maldacena:2013xja,Balasubramanian:2014hda}.

The entanglement spectrum can be extracted from the entanglement R\'enyi entropies $S_n(A) = \frac{1}{1-n} \ln \Tr \rho^n$, if we are able to compute the latter for all real, positive $n$.  These entropies have a further use, in that if we are able to take the limit $n \to 1$, we can compute the von Neumann entropy $S(A) = - \Tr \rho \ln \rho$, which has useful general, thermodynamic-type properties.  The path integral implementation of this limit, as a kind of replica trick, is a powerful technique for computing the von Neumann entropies, particularly for conformal field theories \cite{Calabrese:2004eu,Calabrese:2009qy,Holzhey:1994we}.

We can always write $\rho_A = e^{-2\pi H_A}$ for some ``modular Hamiltonian" $H_A$ acting on $\calh_A$, in which case continuously tuning $n$ is equivalent to tuning the temperature for the system with Hilbert space $\calh_A$ and Hamiltonian $H_A$ in the canonical ensemble.  Extracting the entanglement spectrum from the R\'enyi entropies $S_n(A)$ is the equivalent to transforming to the microcanonical ensemble from the canonical ensemble in this system. This presentation then begs the question of whether there are any phase transitions as a function of $n$.  Such transitions can point to generic features of the entanglement spectrum, perhaps following the discussion in \cite{huller1994first}. In principle they could provide an obstruction to computing the von Neumann entropy if the transition is at $n=1$, although we will point out in the conclusions that this is not necessarily the case. Such transitions have been found, for example, in the $\epsilon$ expansion around $d = 3$ spatial dimensions \cite{Metlitski:2009iyg}, and a characterization of the onset of a class of phase transitions has been made for holographic theories in \cite{Belin:2013dva,Belin:2013uta,Puletti:2017gym}.\footnote{There is also some evidence for a phase transition for the R\'enyi entropies for the entanglement of two intervals with their complement, in 2d CFTS \cite{Belin:2017nze,Dong:2018esp}.}

In this work we will consider the interacting IR fixed point of the $O(N)$ vector model in $d=2$ spatial dimensions, for which $A$ corresponds to the region inside a $2$-dimensional disk of radius $R$.\footnote{Note that we do not perform a singlet projection as opposed to \cite{Klebanov:2002ja}, so the higher-spin dual in that work is not directly relevant to us.}   In this simple case, $H_A$ can be written as the Hamiltonian on $2$-dimensional hyperbolic space $\mH_2$ \cite{Casini:2011a}. In this background, and in the $N\to\infty$ limit, the classic Hohenberg-Mermin-Wagner-Coleman theorem \cite{Mermin:1966,Hohenberg:1967,coleman1973}\ does not immediately pertain \cite{Callan:1989em,Witten:1978qu,Anninos:2010sq}.  We will find that there is an instability in the theory for $H_A$ at $T = \frac{1}{2\pi n} = 1/2\pi$, after the fashion of the instabilities studied in a holographic context \cite{Belin:2013dva,Belin:2013uta}. This instability is towards a symmetry-breaking phase, indicating a possible second-order phase transition there. At present, we cannot completely rule out the possibility there is instead a first-order phase transition at a higher temperature, although we will argue in the conclusions that the results of \cite{Whitsitt:2016irx}\ make this unlikely. That said, the existence of an instability is already interesting and non-trivial, and comprises a non-trivial analytic result for the entanglement structure of an interacting quantum field theory. The story at finite N is also unknown to us, but we discuss some issues and possibilities in Sec.~\ref{Conclusion}.

The outline of our paper will be as follows. In Sec.~\ref{Review}, we review the map between the disk and hyperbolic space in order to relate Re\`nyi entropies and free energies. In Sec.~\ref{ON}, we begin with a review of the interacting $O(N)$ model in flat space, and proceed to study the theory on hyperbolic space $\mH_2$. We find that the theory develops an instability at $T=\frac{1}{2\pi}$, which is our main result. In Sec.~\ref{Conclusion}, we argue that combining our results with the work of \cite{Whitsitt:2016irx}\ points to the transition being second-order, and that this transition at $n=1$ does not invalidate the replica trick calculation.  Wdiscuss the possible physics at finite N, and outline possible future work. The Appendices contain some detailed calculations whose results we use in the main body of the paper.

\section{Review: R\'enyi entropies as free energies} \label{Review}

In this work we consider quantum field theories in $d=2$ spatial dimensions on $\mR^2$. We are interested in the entanglement properties of the ground state, as measured by the reduced density matrix for degrees of freedom inside a circular disc $D_R$ of radius $R$. 

While for general quantum theories,  we can always write $\rho_A = e^{-2\pi H_A}$, there is no guarantee for a general subregion $A$ in a general QFT that the ``modular Hamiltonian" $H_A$ is anything nice, or local.  However, for CFTs in $d$ spatial dimensions, when $\calh_A$ corresponds to the degrees of freedom inside a $d$-dimensional spherical ball of radius $R$, $H_A$ can be written as the Hamiltonian in $d$-dimensional hyperbolic space $\mH_d$, with radius of curvature $R$ \cite{Casini:2011a}.  Note that as defined $H_A$ is dimensionless: we will replace $H_A \to R H_A$ in this formula, so that $\rho = e^{-2\pi R H_A}$ is the canonical ensemble density operator for a theory on hyperbolic space with temperature $T = \frac{1}{2\pi R}$.

The map between the density matrices is induced by a conformal transformation. Here, we recall the conformal transformation from the spherical ball to $\mH_d$ used by  \cite{Casini:2011a} to establish the relationship between density operators. This maps the domain of dependence of the sphericall ball to the hyperbolic space and consequently, relates the modular Hamiltonian on the causal diamond of the disk to the generator of time translations in hyperbolic space. 

One first starts with the flat-space metric
\be
ds^2 = -dt^2+ dr^2 + r^2 d \Om_{d-1}^2
\ee 
Our spherical subregion sits at $r=R, t=0$. We now make the following conformal transformation:
\be \label{conformaltrans}
t = R \frac{\sinh \big(\frac{\T}{R}\big) } {\cosh \R+\cosh \big(\frac{\T}{R}\big)} , \qquad r = R \frac{\sinh \R}{\cosh \R + \cosh \big(\frac{\T}{R}\big)}
\ee 
and obtain the metric on $R \times \mH_d$, but for a conformal pre-factor.
\be \label{hyp_metric}
ds^2 = \bigg(\cosh \R + \cosh \bigg(\frac{\T}{R}\bigg) \bigg)^{-2}[-d\T^2 +R^2 ({d\R}^2+ \sinh^2 \R d \Om^2_{d-1})]
\ee 
where we have the conformal pre-factor in round backets and the quantity inside the square brackets is the metric on $\mH_d \times R$.

Given this identification, 
\begin{enumerate}
\item Equal-time correlation functions on $D_R$ are equivalent, up to conformal factors, to correlators at equal Euclidean time for the theory on $\mH_2\times S^1_{2\pi R}$ -- that is, for equilibrium correlators of the CFT on $\mH_2$ at temperature $T = 1/2\pi R$. 

\item The R\'enyi entropies are
\be
	\frac{1}{1-n} \ln \Tr \rho^n = \frac{2\pi R n}{n-1} F(\beta = 2\pi n R)
\ee
where $F$ is the free energy at inverse temperature $\beta$.
\end{enumerate} 
Our focus will be on the second point: we will use the first point to nail down our renormalization prescription.

For a QFT most of the entanglement between a spatial region and its complement is carried by ultraviolet degrees of freedom near the boundary line/point/surface between them. An explicit calculation on the R\'enyi or von Neumann entropies thus requires a cutoff. In the case at hand, the cutoff and the radius $R$ are the only dimensional parameters.  If we write the cutoff $\Lambda$ in units of momentum, so that the UV corresponds to large $\Lambda$,we expect that  \cite{Metlitski:2009iyg,Liu:2012eea}
\be
	S_n = s_{-1,n} R \Lambda + s_{0,n} + \calo\left(\frac{1}{\Lambda R}\right) + \ldots
\ee
The finite piece $s_0$ is thought to be universal.

Under the conformal transformation, the UV cutoff near the entangling surface becomes an IR cutoff on the volume of $\mH_2$.  We will treat this with a hard radial cutoff following \cite{Casini:2011a,Klebanov:2011uf}; the upshot is that one can replace any (divergent) factors of $Vol(\mH_2)$ with $-2\pi$ \cite{Klebanov:2011uf}.

The leading divergence in the entanglement entropies is physical in continuum QFT (as opposed to theories of quantum gravity, in which case the divergence can be absorbed into the renormalization of Newton's constant).  In addition, any explicit representation of the path integral has additional UV divergences. These are standard QFT divergences, removed from physical quantities via a renormalization prescription for the vacuum energy and the couplings, in order to properly define the path integral in the first place. We will treat them separately, via proper time regularization.

\section{The interacting large-N $O(N)$ vector model}\label{ON}

\subsection{Review: the $O(N)$ model on $\mR^2$.} \label{On_review}

The theory we will focus on is the interacting IR fixed point of the large-N $O(N)$ vector model in $d=2$ spatial dimensions.  We open with a review of this theory on flat Minkowski spacetime $\mR^3$, to introduce the techniques we will use in a more familiar context.  This theory can be written as the infrared fixed point of the theory\footnote{Note that there are several presentations of the underlying physics, corresponding to different UV embeddings. See for example \cite{polyakov1987gauge,ZinnJustin:2003a,Hartnoll:2005yc,sachdev2011quantum}. In this work we use the UV embedding described in \cite{ZinnJustin:2003a,Hartnoll:2005yc}\ and references therein.}
\be
	S = \int d^3x \left[ \half (\pa {\vec\phi})^2 - \frac{h}{4 N} (\vec\phi^2)^2\right]
\ee
where $\vec \phi$ is an $N$-dimensional vector. The infrared critical point corresponds to the limit $h \to \infty$ (with the limit taken through positive values).  We will embed the above Lagrangian in the larger set of theories
\be
	S = \int d^3x \left[ \half (\pa {\vec\phi})^2 - \half m^2 {\vec \phi}^2 - \frac{h}{4 N} (\vec\phi^2)^2\right]
	\label{eq:massivefamily}
\ee
We do this because in our renormalization scheme, quantum fluctuations will push the IR fixed point away from $m^2 = 0$.
Furthermore, as we will see, it is interesting to study the theory as a function of $m^2$.

The theory of \eref{eq:massivefamily} supports a phase transition at zero temperature.
The equilibrium, finite temperature theory can be studied via a Euclidean path integral on $\mR^2\times S^1_{\beta = 1/T}$. At long distances $\delta x \gg \beta$ we may dimensionally reduce the theory to $d = 2$, and the Hohenberg-Mermin-Wagner-Coleman theorem \cite{Mermin:1966,Hohenberg:1967,coleman1973}\ prevents a phase transition in this model.  At large N this implies an infrared singularity in the gap equation. At zero temperature, however, a second-order phase transition can be reached by dialing $m^2$, and corresponds to an ordering transition in which ${\vec \phi}$ acquires a vacuum expectation value. To set ourselves up for a discussion of the theory on $\mH_2$, we review this zero-temperature phase transition.

We start by unpacking the quartic coupling term via a Hubbard-Stratanovich transformation: in Euclidean space, we write 
\be \label{aux_action}
S[\vec{\p},\s] = \int d^3x \sqrt{g} \Bigg(\frac12 (\del \vec{\p})^2  + \frac12 ( m^2+ h \s) \vec{\p}^2 -  \frac{h N}{4}  \s^2 \Bigg)
\ee
Integrating out $\sigma$ in the path integral yields the original action in \eref{eq:massivefamily}.  Note that since the action is quadratic in $\sigma$, this is the same as setting $\s = \frac{\p^2}{N}$, that is, the saddle point value.

Classically, when $m^2 > 0$, the classical vacuum is at ${\vec \phi} = 0$, while for $m^2 < 0$ there is an $S^{N-1}$ family of vacua, with $N-1$ massless Goldstone modes. Quantum effects are important, and spoil this transition at finite temperature.  At $T = 0$, however, there is a transition at a specific value of $m^2$, although quantum effects also push this away from $m^2 = 0$.

To find this point, let us first assume we are in the symmetry-broken phase, for which ${\vec \phi}$ has an expectation value. We will determine the values of $m^2$ for which this is self-consistent.  We can use the $O(N)$ symmetry to rotate ${\vec\phi}$ into the first component:
\be
	\vec{\p}=(\sqrt{N}\p +\dl \p,\pi_{a}), \quad a=1 \cdots N-1
\ee
here $\phi \geq 0$ is the expectation value; $\delta\phi$ the ``Higgs mode" when $\phi \neq 0$; and $\pi_a$ the putative Goldstone modes.  Because the Hubbard-Stratanovich transformation renders the action Gaussian in $\pi$, we can integrate these modes out, to arrive at the partition function
\begin{eqnarray} 
Z[\s] & = & \int [d\s] [d \p] exp \left(- N \int d^3x \Bigg(\half (\del \p)^2 + \frac{(m^2+ h \s)}{2} \p^2 - \frac{h \s^2 }{4} \Bigg) 
\right.\nonumber\\
& & \qquad \qquad \left. + \frac{(N-1)}{2} \Tr \ln(- \del^2 +m^2 + h \s)  \right)\label{full_action}
\end{eqnarray}
The second line is the exponentiated one-loop determinant coming from the Gaussian integral. Note that $m_{eff}^2 = m^2 + h \sigma$ is the effective mass of the putative Goldstone modes $\pi$. Eq. (\ref{full_action}) is a large-N result: the effects of integrating out $\delta \phi$ are subleading in $N$ ({\rm c.f.} \cite{Hartnoll:2005yc}).  

In the $N \to\infty$ limit, the theory can be solved via the saddle point method. The result is the pair of gap equations:
\begin{align} \label{eq:flatsaddle}
& & (-\del^2+m^2 +h \s) \p = 0 \nonumber \\
& & \p^2 - \s + \frac{1}{V}\Tr \Bigg(\frac{1}{-\del^2+ m^2 + h \s} \Bigg) =0
\end{align}
where $V$ is the regularized volume of $\mR^3$. The trace can be computed explicitly:
\begin{eqnarray}
	\Tr \Bigg(\frac{1}{-\del^2+ m^2 +h \s} \Bigg) & = & V \int_{\Lambda^{-2}}^{\infty} ds \int \frac{d^3 p}{(2\pi)^{3/2}} e^{- s (p^2 + m^2 + h \sigma)}\nonumber\\
	& = & \frac{V}{2^{3/2}} \int_{\Lambda^{-2}}^{\infty} \frac{ds}{s^{3/2}} e^{- s(m^2 + h \sigma)}	\label{eq:flattrace}
\end{eqnarray}

We assume that Lorentz invariance will not be broken, in which case we can search for solutions to these equations with constant $\p,\s$.  In this case the first equation in \eref{eq:flatsaddle} has two branches.  In one branch, $\phi = 0$ and $m_{eff}^2 = m^2 + h \sigma$ is unconstrained, although stability of the theory demands that $m_{eff}^2 \geq 0$ for the minimum of the free energy. In the other branch, $\phi \neq 0$ and $m^2 + h \sigma = 0$. As we started, self-consistently, with the assumption $\phi > 0$ we will look for solutions in this class. Since $m^2 + h\sigma = 0$, the trace is simple to compute, 
\be
	\frac{1}{V} \Tr \Bigg(\frac{1}{-\del^2+ m^2 +h \s} \Bigg) = \frac{\Lambda}{\sqrt{2}} 
\ee
Thus, there is a solution for
\be
	\phi^2 = \sigma - \frac{\Lambda}{\sqrt{2}} = - \frac{m^2}{h} - \frac{\Lambda}{\sqrt{2}}
\ee
This has a nonzero solution when $m^2 < m_c^2 = -  h \frac{\Lambda}{\sqrt{2}}$.  

At $m^2 = m_c^2$, $\phi^2 = 0$ and $m_{eff}^2 = 0$.  At long distances, or equivalently when $h \to\infty$ for fixed distances, this describes the conformal point of the model, the interacting $O(N)$ CFT at large N.

For $m^2 > m_c^2$, we must set $\phi^2 = 0$. The fluctuations are now $O(N)$ symmetric, and the remaining action is given in 
\eref{full_action} with $N-1$ replaced by $N$. At large $N$, the saddle point equations are still \eref{eq:flatsaddle}, with the effective mass $m^2 + h\sigma > 0$. Performing the proper time integral in \eref{eq:flattrace} and discarding terms which vanish as $\Lambda \to \infty$, the saddle point equation becomes:
\be
	\sigma = \frac{\Lambda}{\sqrt{2}} - \sqrt{\frac{\pi}{2}} \sqrt{m^2 + h \sigma}
\ee
We can simplify this by defining $\sigma = \frac{\Lambda}{\sqrt{2}} + \delta\sigma$, and $m^2 = m_c^2 + \delta m^2$. Squaring both sides,
\be
	\delta \sigma^2 = \frac{\pi}{2} \Big( \delta m^2 + h\delta\sigma \Big)
\ee
which always has a real solution for $\delta \sigma$; the negative branch $\delta \sigma = \frac14(\pi h - \sqrt{ \pi^2 h^2 + 8 \pi \delta m^2})$ vanishes as $\delta m^2 \to 0$, as it should.

\subsection{The $O(N)$ model on $\mH_2$: setup} \label{On_hyperbolic}

We now pass to the system of interest: the $D=3$-dimensional, Euclidean $O(N)$ model on the manifold $\mathbb{H}_2 \times S^1 $, where $\mathbb{H}_2$ is the two-dimensional hyperbolic space with metric
\be \label{eq:h2metric}
{ds}^2 = R^2({d\R}^2 + \sinh^2{\R} {d\p}^2) 
\ee 
where $\R \in [0,\infty)$ and $\p \in [0,2\pi)$. Here $R$ is the radius of $\mH_2$, and is equal to the radius of the disc whose reduced density matrix we are trying to compute.

We work with the action:
\be \label{org_action}
S[\vec{\p}] = \int d^3x \sqrt{g} \Bigg(\frac12 (\del \vec{\p})^2  + \frac12 \left(m^2  + \frac{1}{8} {\cal R}\right)\vec{\p}^2 + \frac{h}{2N} (\vec{\p}^2)^2 \Bigg)
\ee
where $g$ is the determinant of the metric on $\mathbb{H}_2 \times S^1 $, ${\cal R}$ the Ricci scalar, and the indices are contracted using the curved space metric.  We have included an explicit curvature coupling appropriate for a conformally coupled scalar.  For this spacetime, it is constant (and negative), and could be absorbed into $m^2$: we often simply use the combination $M^2 = m^2 + \frac{1}{8} {\cal R}$. As before, this is a theory with a dimension-1 coupling $h$. In the present case we can form a dimensionless coupling $h R$ or $h \beta$; the IR corresponds to $h \to\infty$ at fixed $R,\beta$ or to $R,\beta$, and all distances probed scaled to infinity at fixed $h$. Note that the family of QFTs labelled by $h$ are not, for general $h$, simply related to the same family in flat space,
as the coupling is not exactly marginal.  We are nonetheless assuming that the IR limit $h \to \infty$ corresponds to the interacting $O(N)$ model on $\mH_2\times S^1$. This is manifestly true at distances smaller than $R$, $\beta$; we are thus assuming that the effects of the background curvature and topology do not drive \eref{org_action} to a different fixed point.

Again, we use the Hubbard-Stratonovich trick to rewrite the action using an auxiliary field $\s$ which is also integrated over:
\be \label{aux_action}
S[\vec{\p},\s] = \int d^3x \sqrt{g} \Bigg(\frac12 (\del \vec{\p})^2  + \frac12 ( M^2+ h \s) \vec{\p}^2 -  \frac{h N}{4}  \s^2 \Bigg)
\ee

Our goal is to understand the phase structure of the IR CFT as a function of temperature. We will do this, as before, by studying the saddle point equations as $N \to \infty$ after integrating out fluctuations of the scalars ${\vec \phi}$:
\begin{align}\label{gap_eq}
(-\nabla_{\mathcal{H}}^2+M^2 + h \s) \p =0 	\nonumber	\\
\p^2 -\s + \frac{1}{V} \Tr \Bigg(\frac{1}{-\nabla_{\mathcal{H}}^2+ M^2 + h \s} \Bigg) =0 
\end{align} 
(up to terms of order ${\cal O}(1/N)$, as before), where $V$ is the regularized volume of $\mH_2 \times S^1$, and ${\cal H} = \mH_2\times S^1$. To solve these, we need to choose the proper boundary conditions for $\sigma,\phi$, appropriate for the R\'enyi entropy calculation. Furthermore, the trace contains UV divergences, and we will find that we must select the counterterm for the mass of ${\vec \phi}$ to get the desired IR physics. We now turn to each of these issues.

\subsection{Boundary conditions on $\mH_2$}

If we were simply quantizing a {\it free}\ scalar field on $\mH_2$, we would choose boundary conditions so that the modes are normalizable. Such modes will fall off exponentially in $\rho$. This should also be the correct choice for the R\'enyi entropy. One argument starts with the observation in \cite{Agon:2013iva}, that the reduced density matrix on the disc for a free scalar field in its ground state breaks up into superselection sectors, each of which corresponds to a different Dirichlet condition on the scalar. A Dirichlet condition on the boundary of the disc becomes, under the conformal transformation to $\mH_2\times R$, exponential falloff.  The boundary value of the scalar maps to the amplitude of the decaying mode, with the decay rate being governed by the conformal transformation. In particular, the constraint $\phi(|\vec{x}|=R,t=0)=\phi_0$ for a free field on the boundary of a disk, of radius $R$, becomes $\phi(\rho,\tau = 0) = \frac{1}{\sqrt{\cosh\R+1}} \phi_0$, as $\rho \to \infty$,\ on $\mH_2 \times R$ after making the conformal transformation in \eref{conformaltrans}. A constant field mode is not in the Hilbert space of the free theory. We will also not allow it as a saddle point. The reader may object that this is also true in flat space, for which expectation values of a scalar field label superselection sectors, yet in studying statistical mechanics one may vary the constant mode. For the free conformal scalar in hyperbolic space this would run us into trouble, as an arbitrarily large constant mode would lower the free energy indefinitely, due to the negative $m^2$ from the conformal coupling to negative curvature. A direct computation of the free energy reveals no such pathology \cite{Klebanov:2011uf}, so we simply forbid non-zero constant values of ${\vec \phi}$.

We therefore expect that for the interacting theory, a constant mode for ${\vec \phi}$ should also not be an allowed saddle point.\footnote{See \cite{Belin:2013dva}\ for a related discussion in the context of holography.} On the other hand, we believe that $\sigma$ can have a constant expectation value. Recall that $\sigma = {\vec\phi}^2$ is the saddle point equation of motion for $\sigma$, before integrating out ${\vec \phi}$.  In the free theory it is perfectly possible for ${\vec\phi}^2$ to have a constant expectation value at finite temperature even if $\phi$ falls off exponentially. This was shown via path integral techniques in \cite{Herzog:2016aa}; in Appendix \ref{free_thermalonepoint}, we show this directly using the operator approach, in order to see how thermally excited exponentially decaying modes can combine to form a constant value of ${\vec \phi}^2$.

\subsection{Renormalization}

The goal of our calculation is to study the phase structure of the interacting ${\cal O}(N)$ CFT on $\mH_2\times S^1_{\beta}$, as a function of $\beta$. This is based on the identification in Sec.~\ref{Review}, which includes the identification between equal time correlators on the disc and thermal correlators on  $\mH_2\times S^1_{R}$.  In the discussion above we are working with a non-conformal theory, under the assumption that the desired behavior emerges in the IR limit $h \to \infty$. Note that away from the UV or IR fixed points, \eref{org_action} is not invariant under conformal transformations.  We will assume that the limits $h \to 0,\infty$ match;  in particular the correlators at equal Euclidean time should map into each other under conformal transformations, in this limit \cite{Casini:2011a}. Note that the two theories {\it do}\ match for distances $\ll R$, so we are making an assumption that the spacetime curvature is not introducing a new fixed point.

Starting with this assumption, we must fix the the bare mass $m^2$ -- and thus the effective mass $M'^2$ -- by a renormalization prescription, to ensure that we hit the correct IR fixed point. The identification of this prescription was, at least to us, nontrivial.  $M'^2 = m^2 + h \s + \frac{1}{8}{\cal R}$ is the effective mass for the Goldstone modes in the symmetry-broken phase, and of all of the scalar fluctuations in the symmetry-unbroken phase.  But, the saddle point solution for $\s$ depends on the temperature. One might have stated that the modular Hamiltonian is meant to be the CFT Hamiltonian  on $\mH_2$, in which case it would make sense to demand that the effective mass $M'^2$ takes the conformal value $\frac{1}{8} {\cal R}$ at zero temperature. Instead, our guide will be the fact that ground state equal-time correlation functions in the disc map to thermal correlators on $\mH_2$ for $\beta = R$ under the conformal transformation in \cite{Casini:2011a}.  We use this fact to nail down the correct value of $m^2$.  We focus on correlators of ${\cal O}_2 = \frac{1}{\sqrt{N}}{\vec \phi}^2$, a conformal primary with operator dimension $2$ in the interacting fixed point, and dimension $1$ in the free $h = 0$ theory. 

Our starting point is the following observation. We can consider our interacting CFT as the IR fixed point of the free theory perturbed by the double-trace operator $ h {\cal O}_2^2$.  Let us write the two-point function $F$ of ${\cal O}_2$ in the UV theory as $F(x,y) = \bra{x} {\widehat F}\ket{y}$, where $\ket{x}$ is the position eigenstate for a quantum-mechanical particle on $\mH_2\times S^1$, and ${\hat F}$ an operator on that space. In the free theory, ${\vec \phi}$ are conformally coupled scalars on $\mH_2\times S^1_R$, and we can use Wick's theorem to write
\be   	\label{fUV}
F(x,y) = \left( \bra{x} \frac{1}{- \del^2 - \frac{1}{4 R^2}} \ket{y} \right)^2
\ee 
where $\del^2$ is the Laplacian on $\mH_2\times S^1_R$, and the appearance of $-1/(4R^2)$ is the result of the conformal coupling to the background curvature. It is straightforward to show that \eref{fUV} is equivalent to the result one would get from a conformal transformation from correlators on the disc $D_R$ in an equal-time slice of $\mR^3$.

The IR two-point function can be written as  $G = \bra{x} {\widehat G}\ket{y}$. We can run the arguments in Section 2\ of \cite{Gubser:2003}, through equation (7), in any background spacetime, to find:
\be
	{\widehat G} = \frac{{\widehat F}}{1 + h {\widehat F}}\label{eq:Gg}
\ee
for any $h$. In the limit $h \to \infty$, $ h{\widehat F} \gg 1$ for any matrix elements of ${\widehat F}$ in spectral or position space, the correlators can be written as:
\be
	{\widehat G}_{IR} \sim \frac{1}{h} - \frac{1}{h^2 {\widehat F}}\label{eq:ircorrident}
\ee
In flat $\mR^3$, the second term in \eref{eq:ircorrident} $G_{IR}$ is the correlator of ${\cal O}_2$ in the interacting CFT \cite{Gubser:2003}, with the first being a contact term. We will assume that, on $\mH_2\times S^1_R$, when $F(x,y)$ is the UV correlator in \eref{fUV}, ${\widehat G} = - 1/(h^2 {\hat F})$ is the correlator of the IR fixed point.  

To ensure this, we compute $G_{IR}$ directly and explicitly in our model, using large-N techniques, and fix the renormalization prescription to ensure that it is equal to the right hand side of \eref{eq:ircorrident}. We consider the action \eref{aux_action} after 
rescaling $\s  \rightarrow \s/(h\sqrt{N})$ and including a coupling $J {\cal O}_2 = J\frac{{\vec \phi}^2}{\sqrt{N}}$, in order to match the discussion in \cite{Gubser:2003} and thus the equations \eref{eq:Gg}, \eref{eq:ircorrident}. Integrating over ${\vec \phi}$, we find:
\be \label{largeN_action}
S[\s] = \frac{N}{2} \Tr \ln\Bigg(-\del^2+M^2+\frac{\s+J}{\sqrt{N}}\Bigg)-\int d^3x \sqrt{g} \frac{1}{4h}  {\s} ^2\ .
\ee 
Next we write $\s$ as a sum of its (constant) divergent part $\s_0$, which solves the gap equation when $J = 0$, and a finite part $\delta \s$, and  expand \eref{largeN_action} to keep only terms which are ${\cal O}(1)$ or larger in the large-$N$ expansion, to find:
\begin{eqnarray} \label{eq:lN_quad}
S[\s]& =& f(\s_0)+\frac{\sqrt{N}}{2V} \Tr\Bigg(\frac{1}{-\del^2+M'^2}\Bigg)J-\frac{1}{2V} \Tr\Bigg(\frac{1}{-\del^2+M'^2}\Bigg)^2 (\delta \s+J)^2
\nonumber\\
& & \qquad\qquad + \int d^3x \sqrt{g} \Bigg( -
\frac{\delta \s^2}{4h}\Bigg)
\end{eqnarray}
where $f(\s_0)=-\frac{V \s_0^2}{4 h}+\frac{N}{2} \Tr \ln(\del^2+M'^2)$ are the $J$-independent terms, and $V$ is the (regularized) volume of $\mH_2 \times S^1$. 

In the limit $h \to 0$, corresponding to the UV, the saddle point for $\delta \sigma$ is $\delta \sigma = 0$; inserting this in \eref{eq:lN_quad}, we find that the two-point function of ${\cal O}_2$ is
\be
	{\widetilde F} = \left(\bra{x} \frac{1}{-\del^2 + (M')^2}\ket{y}\right)^2 \equiv \bra{x}{{\hat{\tilde F}}}\ket{y}
\ee
where the tilde denotes the computation using our large-N action. In the IR limit $h \to\infty$, the two-point function is 
\be
\widehat{{\widetilde G}} =  \frac{\widehat{\widetilde{F}}}{1+h {\widehat{\widetilde{F}}}}
\ee 
Again, for distance scales such that $h {\widehat{\widetilde F}} >>1$ (where we understand ${\widehat{\widetilde F}}$ to correspond to matrix elements in some appropriate basis), we find
\be
{\widehat{\widetilde G}} = \frac{1}{h} -\frac{1}{h^2 {\widehat{\widetilde F}}}+ \cdots
\ee
This matches \eref{eq:Gg}, \eref{eq:ircorrident} if we set $(M')^2 = - \frac{1}{4R^2}$ at $\beta = R$, which we will do below.

\subsection{Solving the gap equation near $\beta = R$} 

We now focus our attention on the gap equations in \eref{gap_eq}. In the case that $\sigma$ is constant, we can write the trace explicitly, using the results in Appendix \ref{saddlepoint_int}:
\begin{eqnarray} \label{gap_int}
& & (-\nabla_{\mathcal{H}}^2+M^2 +h \s) \p =0 \nonumber \\
& & \p^2 - \s + \frac{1}{(2\pi)R^2\bt} \int_{\ep}^{\infty} ds \sum_{\ell} \int_{0}^{\infty} d \la_0 \la_0 \tanh(\pi \la_0 ) e^{-s \Big( \frac{\ell^2}{\bt^2}+\frac{\la^2}{R^2}+M^2+ h \s \Big)}=0\nonumber\\
\end{eqnarray}
where we have regularized the divergent $\la_0$ integral with a cutoff $\ep= \frac{1}{\La^2}$.  In these equations, $\phi = 0$ is aways a solution. Solutions with $\phi \neq 0$ may exist, but will be nonconstant on $\mH_2$, and will require nonconstant $\sigma$: the trace in \eref{gap_eq} is then considerably more challenging to compute. We will focus on the simpler question of when the $\phi = 0$ solution is a local saddle point or a local minimum, indicating a possible phase transition, as in the former case there are directions in field space which will continuously lower the free energy. As we will see, once we fix $m^2$ with our renormalization prescription, $\beta = R$ is the boundary in temperature between stable and unstable free energy at $\phi = 0$.

At $\beta = R$, we fix $M'^2 = M^2 + h \sigma = - \frac{1}{4R^2}$. In this case the integral in the second line of \eref{gap_int} can be evaluated analytically (see Appendix \ref{saddlepoint_int}), yielding the saddle point solution
\be
\s = \alpha \La , \qquad \alpha =\frac{\sqrt{\pi}} {(2\pi)}
\ee 
This fixes $M^2 = - \alpha\Lambda - \frac{1}{4 R^2}$. Note that there is no IR divergence in this calculation, even at the conformal point at which the eigenvalue spectrum of the kinetic operator is continuous down to $\lambda = 0$. The essential point is that the infrared behavior of scalars in $\mH_2$ is weaker than in $\mR^2$ \cite{Callan:1989em}, and does not in and of itself give an obstruction to a finite-temperature phase transition.

The trace in the gap equation (\ref{gap_eq})\ is difficult to compute explicity for general $\beta \neq R$, even for constant $\sigma$.  We consider small perturbations in $\beta$ around $\beta = R$.  In Appendix \ref{saddlepoint_int}, we also show that to linear order in $\delta\beta = \beta - R$, the saddle point equations at $\phi = 0$ yield $(M')^2 + \frac{1}{4 R^2} = - c \delta\beta$, where $c$ is a positive real number. Using this fact, we can show that there is an instability that sets in for $\beta > R$.

For $\beta > R$ or $\delta\beta > 0$, we can expand the action \eref{aux_action} about the saddle point solution
$\phi = 0$, $M^2 + h \s = - \frac{1}{4 R^2} - c\delta \beta$.  Integrating by parts, the action for $\delta{\vec \phi}$ takes the form
\be
	S_{\delta\phi} = \int d^3 x \delta{\vec\phi}\cdot\left( - \nabla_{\mH_2}^2 - \partial^2_{S^1} - \frac{1}{4 R^2} - c \delta \beta\right)\delta{\vec \phi}
\ee
The essential point is that with our boundary conditions, the operator $- \nabla_{\mH_2}^2 - \frac{1}{4 R^2} - \partial^2_{S^1}$ has eigenvalues which are positive, continuous, and extend to zero, as is implied by \eref{hyp_laplaceeqn}.  Thus, the kinetic operator for $\delta{\vec \phi}$ will have negative eigenvalues bounded by $-c\delta\beta$ when $\delta \beta > 0$, and so there are directions in field space which will lower the free energy.

By a similar token, for $\delta \beta < 0$, the action quadratic in $\delta{\vec\phi}$ is strictly positive about the ${\vec \phi} = 0$ saddle point. We have not been able to show whether there is or is not another local minimum for non-zero ${\vec\phi}$ in this range, much less what the free energy might be.  The solutions to \eref{gap_eq} require non-zero $\phi$ and thus non-zero $\sigma$. Without finding or ruling out a solution, we cannot at this time say whether there is a first or second-order transition as a function of temperature in this model. We close by noting that this treatment works strictly in the $N\to\infty$ limit, for which the saddle point equations in $\phi,\sigma$ capture the theory completely.

\section{ Conclusions} \label{Conclusion}

We have shown that with a few plausible conjectures, there is an instability in the large-N $O(N)$ model on $\mH_2\times S^1_{\beta}$ at $\beta = R$, indicating a phase transition in the R\'enyi entropy for the reduced density matrix describing the spatial region $D_R$ of the same theory in $\mR^3$, in the vacuum state. A number of questions arise:
\begin{enumerate}
\item One conjecture we rely on, which is commonly made for the $O(N)$ model on other spacetimes ({\it e.g.} \cite{Hartnoll:2005yc,Klebanov:2011aa}), that the IR fixed point of \eref{org_action} at $\beta = R$, with $m^2$ chosen according to our renormalization prescription, is in fact conformal to the IR fixed point of the same action on the domain of dependence of $D_R$ in $\mR^3$.  It is certainly true that we will get the correct behavior at distances $\ll R$, so it is a question of whether the spacetime curvature might generate a fixed point distinct from the one we wish to reach. Further study along the lines of \cite{Osborn:1991gm,Dong:2012ua}\ would be of interest.
\item We have made an assumption about the boundary conditions for our model based on those for the UV theory, despite the fact that the RG trajectories will not completely match up. This would be worth checking.
\item We have found an instability for the theory at $\beta = R$. What we do not know is whether this corresponds to a second-order phase transition, or rather indicates a kind of spinodal point of a first-order transition, as appears to be the case with the Hagedorn transition in string theory \cite{Atick:1988si,Kruczenski:2005pj}.  Note that the same question pertains to the holographic discussion of R\'enyi phase transitions in \cite{Belin:2013dva,Belin:2013uta}. It is notable that the extension of calculations of \cite{Belin:2013dva}\ to bulk theories with higher-curvature corrections to Einstein gravity \cite{Puletti:2017gym}\ exhibits first-order transitions in the R\'enyi entropies of putative duals (at $n > 1$).  
\item To understand this, one would need to solve (at least in some controlled approximation) the gap equation \eref{gap_eq} for non-zero values of ${\vec \phi}$.  This will be more challenging than the case of flat space, as the boundary conditions require that nontrivial solutions be nonconstant (as also pointed out in \cite{Belin:2013dva}).
\item Finally, we would like to better understand finite-N corrections.  In other contexts \cite{Witten:1978qu,Anninos:2010sq}, for flat spatial dimensions in the thermodynamic limit, finite-N effects can change the nature of the phase transition to one of Berezinski-Kosterlitz-Thouless type \cite{Berezinsky:1970fr,Kosterlitz:1973xp}. Hyperbolic space tames the IR divergences that would appear in flat space, and can change the location of phase transitions \cite{Callan:1989em}. An additional complication here is that the potential ordering is not via constant expectation values for the scalar, but via normalizable modes on $\mH_2$, for which fluctuations between different directions in the field space $\mR^n$ have finite action.  
\end{enumerate}

We close with a comment on the validity of the replica trick and with an argument that the transition at hand is indeed second order. These comments are motivated by the replica trick calculation of \cite{Whitsitt:2016irx}, which successfully reproduces the direct calculation of the entanglement entropy of \cite{Klebanov:2011aa}. One might worry that this result is in tension with our claim of a transition at $n=1$.

However, this is not necessarily the case. Let us imagine first that there was a first-order phase transition at $n=1$. This means that there would be a kink in the free energy. At large $N$, this occurs because two local minima of the free energy exchange dominance. If we  expand the free energy along either branch in a power series on $(n-1)$ it will be analytic, and the $n \to 1$ limit still gives the correct free energy. In the calculation of \cite{Whitsitt:2016irx}, this could happen if the function $f$ in equation (16) changes values as $n-1$ changes sign. But that term drops out of the entanglement entropy calculation.

If the transition is second order, this is still not an obstruction to the use of the replica trick in \cite{Whitsitt:2016irx}. That work relies on the first two terms $F \sim F_0 + (n-1) F_1 + \ldots$, where the further terms vanish faster than linearly. The free energy for a system with a second order transition could have the expansion
\be
	F = f(n-1) + (n-1)^a g(n-1) + \ldots
\ee
where $f,g$ are analytic functions of $n-1$, and $a > 1$ is noninteger. As $n\to 1$, the constant and linear term would dominate the calculation and the replica trick would still give the right answer.

Furthermore, the fact that the replica trick calculation \cite{Whitsitt:2016irx}\ gets the right answer at leading order in $1/N$ provides a good argument that there is not a first-order transition at $n > 1$, as this would occur via a different solution to the gap equations which would have lower free energy, and the free energy in \cite{Whitsitt:2016irx}\ would give the wrong result.  In principle there could be a first-order transition at $n=1$, but a distant saddle point becoming dominant at $n=1$ just as the $\phi = 0$ saddle becomes unstable is a highly nongeneric (if not completely ruled out) situation.

All this being said, the final two questions listed above are important for understanding the implications of the instability we found here for the the structure of the entanglement spectrum. 

%
%



\paragraph{Acknowledgments: } This work grew out of preliminary discussions and calculations done with Howard Schnitzer and Ida Zadeh, and we would like to thank them especially. We would also like to thank Alastair Grant-Stuart, Jonathan Harper, Aram Harrow, Matthew Headrick, Alex Maloney, Aditi Mitra, Mark Mueller, Matthew Roberts, Andrew Rolph, and Bogdan Stoica for helpful discussions. We would like to thank Alexandre Belin, Alex Maloney, Seth Whitsitt, and William Witczak-Krempa for very helpful correspondence following the posting of the first version of this paper on the arxiv. The work of HRH and AL was supported in part by the Department of Energy under DOE grant de-sc0009987. The work of HRH was also supported by the Simons Foundation as part of the ``It from Qubit" grant. The work of SS was supported by the Department of Energy under contract DE-SC0010010 Task A.

\appendix
\section{Thermal expectation value of $:{\vec \phi}^2:$ on $\mathcal{H}$} \label{free_thermalonepoint}

 Let us consider a conformally-coupled free-scalar field on $\mathcal{H}$ defined by the Euclidean action:
 \be
 S[\p] = \int d^3 \bar{x} \Bigg[ -\frac12 \sqrt{g} \Bigg(  g^{\mu \nu} \nabla_{\mu} \p \nabla_{\nu} \p + m'^2 \p^2 +\frac18 R^{(3)} \p^2 \Bigg) \Bigg]
 \ee
  The equation of motion are given by,
  \be
  \Bigg(\nabla^2  -m'^2 -\frac{R^{(3)}} {8} \Bigg) \p =0
  \ee 
  In order to quantize the scalar field, let us re-cast the above equation as a wave equation:
  \be
  \Bigg(\del_{t}^2 - \frac{\la^2}{R_H^2} - M^2 \Bigg) \p =0 
  \ee
  where the constant Ricci scalar has been absorbed in the mass term.
 We now define the frequencies $\omega_{\la_0}^2= \frac{\la^2}{R_H^2} +M^2 $ (recall that $\la^2 =\la_0^2+\frac14$) and using the eigen basis \ref{basis_hyp}, express the scalar in terms of creation and annihilation operators:
 \be \label{psi_modes}
 \p(x) = \int d \la_0 \sum_{n} \frac{1}{\sqrt{2 \omega_{\la_0} }} \mu_n(\la_0) \Big[ a_{n,\la_0} \theta_{n,\la_0} (\R, \psi) +  a^{\dagger}_{n,\la_0} (t) \theta^{*}_{n,\la_0} (\R, \psi) \Big]
 \ee
 where $\theta_{n,\la_0} (\R,\p)= e^{i n \psi} P^{|n|}_{-\frac12 + i \la_0} (\cosh\R)$, and the density of Laplacian eigenstates $\mu_n(\la_0)$ is given by
\be
\label{eq:muo}
\mu_n(\la_0) = \frac{1}{2\pi} \frac{ |\Gamma (\frac12 -|n| + i \la_0)|^2}{|\Gamma(i \la_0)|^2} = \frac{\la_0 \tanh(\pi \la_0)}{2 \pi} \frac{1}{|(\frac12 + i \la_0) \cdots (i \la_0 +|n| -\frac12)|^2} \\
\ee
 The conjugate field $\pi(x,t)$ becomes,
 \be \label{pi_modes}
 \pi(x) = \int d \la_0 \sum_{n} \frac{-\sqrt{ \omega_{\la_0}} }{\sqrt{2 }} \mu_n(\la_0) \Big[ a_{n,\la_0}  \theta_{n,\la_0} (\R, \psi) -  a^{\dagger}_{n,\la_0} \theta^{*}_{n,\la_0} (\R, \psi) \Big]
 \ee
 Now, it is easy to check that imposing the commutation relations 
 \be
 [a_{n,\la_0} (t),a^{\dagger}_{n',\la_0'} (t)] = \frac{ \delta_{nn'} \delta(\la_0 -\la_0')} {\mu_n(\la_0)}
 \ee
  on the creation and annihilation operators is equivalent to imposing the following canonical commutaion relations on the fields
  %
  \be
  [\p(x,t),\pi(x',t)] = \frac{\delta(\psi -\psi') \delta(\la_0 -\la_0')}{R_H^2 \sinh(\R)}
  \ee
  %
  
%
  
  With these results, we wish to show that composite operators can have constant expectation values on $\mH_2$, even if the boundary 	conditions on ${\vec \phi}$ forbid it.  Let us consider the operator $\phi^2$.  We normal-order to ensure that the two-point function vanishes:
  \begin{eqnarray}
  \langle 0 | :\p^2(x,t): |0 \rangle & = & \langle 0| \int d \la_0 d \la_0' \sum_{nn'} \frac{\mu_{n'}(\la_0') \mu_n(\la_0)} {\sqrt{\omega_{\la_0}\omega_{\la_0'}}} a^{\dagger}_{n',\la_0'} a_{n,\la_0} \theta_{n,\la_0}(\R,\psi) \theta^{*}_{n,\la_0'}(\R,\psi)| 0 \rangle\nonumber\\
  & = & 0
  \end{eqnarray}
  Here, $|0 \rangle $ refers to the vacuum on $\mH_2 \times R$. We now make the system thermal by making the time dimension compact 
  \be \label{twopoint_thermal}
    \langle :\p^2(x): \rangle_{\beta} = \frac{Tr( e^{-\beta H} :\p^2(x): )} {Tr(e^{-\beta H} )}
  \ee
  where H is the modular Hamiltonian given by:
  \be \label{modularH}
  H(\pi,\p) = \int d^2x \frac{\sqrt{g} }{2} \Big[ \pi^2 + (\del_i \p)^2 +m^2 \p^2 \Big]
  \ee

 In order to compute \eref{twopoint_thermal}, we need to express the modular Hamiltonian \eref{modularH} in terms of the creation and annihilation operators using \eref{psi_modes}, \eref{pi_modes}. Using the completeness relation for the $\theta$s, we find the following normal-ordered Hamiltonian
 \be
 : H: = \int d \la_0 \sum_n  \mu_n(\la_0) \omega_{\la_0} a^{\dagger}_{n,\la_0} a_{n,\la_0}
 \ee 
We can see that the energy eigenvalues $\omega_{\lambda_0}$ are independent of $n$. The thermal expectation value of $:\phi^2(x):$ is found to be
\be
\langle :\phi^2(x): \rangle_{\beta} = \int \la_0 \sum_{n} \mu_{n}(\la_0)
\frac{1}{e^{\beta \omega_{\la_0}} - 1}
\langle n,\la_0 | :\phi^2(x):  | n,\la_0 \rangle
\ee
where the exponential pre-factor arises because we are working with a boson. After a bit of algebra, this expression becomes
 \be
\langle :\phi^2(x): \rangle_{\beta} =  \int_{0}^{\infty} \frac{d \la _0}{2 \omega_{\la_0}} \sum_{n} \mu_{n} (\la_0) 
|\theta_{\la_0,n}(x)|^2
\frac{1}{e^{\beta \omega_{\la_0}} - 1}
 \ee
At this point, it is useful to use the identity \cite{Balazs:1986uj}:
\be 
\sum_{n= -\infty}^{\infty} \frac{|P^{|n|}_{-\frac12+i \la_0}(\cosh(\R)) |^2} {|(\frac12 + i \la_0) \cdots (i \la_0 +|n| -\frac12)|^2} =1
\ee
which, together with Eq~(\ref{eq:muo}) brings the expectation value to the form
\be
\langle :\phi^2(x): \rangle_{\beta} =  \int_{0}^{\infty} \frac{d \la _0}{2 \omega_{\la_0}} \sum_{n} \frac{\la_0 \tanh(\pi \la_0)}{2 \pi}
\frac{1}{e^{\beta \omega_{\la_0}} - 1}
\ee
which is manifestly constant.
 
 
\section{The saddle-point integral} \label{saddlepoint_int}
In this section, we analyze the gap equations of \eref{gap_eq} on $\mH_2 \times S^1$ in detail. Before we get into this discussion, it is useful to briefly review the Laplacian and its eigenfunctions on hyperbolic space. We start with the Laplacian on $\mathcal{H}$ given by $\nabla^2_{\mathcal{H}}$:
\be \label{hyp_laplacian}
\nabla^2_{\mathcal{H}} = \frac{1}{R^2} \Bigg[ \del_{\R}^2+ coth(\R)\del_{\R}+ \frac{1}{\sinh^2(\R)} \del_{\p}^2   \Bigg] + \frac{1}{\bt^2} \del_{t}^2
\ee
Here, $\bt$ is the radius of the thermal circle and $R$, the radius of $\mH_2$ and the variables $\R,t$ and $\p$ are the embedding coordinates in the metric of \eref{eq:h2metric}. The corresponding eigenfunctions are,
\be \label{basis_hyp}
\langle \la n \ell | \R \p t \rangle = \Ph_{\la_0,n,\ell} (t,\p,\R) = e^{i\ell t} e^{i n \p} P^{|n|}_{-\frac12 + i \la_0} (\cosh\R)
\ee  
where $P^{|n|}_{-\frac12 + i \la_0} (\cosh\R)$ are associated Legendre functions \cite{Balazs:1986uj}.
The eigenfunctions satisfy the equation:
\be \label{hyp_laplaceeqn}
\nabla^2_{\mathcal{H}} \Ph_{\la_0,n,\ell}  = - \Bigg(\frac{\la^2}{R^2} + \frac{\ell^2}{\bt^2} \Bigg)  \Ph_{\la,n,\ell}   ; \qquad \la^2 = \la_0^2 + \frac14
\ee
%
Here, $\la$ is the eigenvalue of the Laplacian on the the hyperbolic space and $\la_0$ is the associated wave number. Thus, the eigenspectrum of the hyperbolic Laplacian is $\la^2 \in \big[\frac14, \infty\big)$ \cite{Balazs:1986uj}. 

The position eigenkets are normalized as follows:
\be
\langle \bar{x} |\bar{y} \rangle= \langle \R \p t | \R' \p' t' \rangle = \frac{\dl(\R- \R')\delta(\p- \p')\delta(t- t')}{\bt R^2 \sinh(\R)}
\ee
 We now have all the ingredients to evaluate the trace term in \eref{gap_int}. Using the eigenbasis \eref{basis_hyp} to express the trace, we get:
 \be \label{gap_int2}
 I(\ep)= \frac{1}{(2\pi)R^2\bt} \int_{\ep=\La^{-2}}^{\infty} ds \sum_{\ell} \int_{0}^{\infty} d \la_0 \la_0 \tanh(\pi \la_0 ) e^{-s \Big( \frac{\ell^2}{\bt^2}+\frac{\la^2}{R^2} + M'^2 \Big)}
 \ee
 where we have introduced a cutoff $\La$ and $M^2+ h \s =M'^2$ is the effective mass. 
 
 
Let us evaluate the integral \eref{gap_int2} at the conformal point $R=\bt$. We set the effective mass to its conformal value $M'^2=-\frac{1}{4R^2}$ to find:
 \begin{align} \label{gap_int3}
 I(\ep) & = \frac{1}{2\pi R} \int_{\ep=(\La R)^{-2}}^{\infty} ds \sum_{\ell} \int_{0}^{\infty} d \la_0 \la_0 \tanh(\pi \la_0 ) e^{-s(\ell^2+\la_0^2)} \\
 &= c_{-1} \La + \frac{c_0}{R} +\frac{1}{R} \sum_{n \geq 1} c_n \ep^{n/2} 
 \end{align}
 where we have scaled the variable $s \rightarrow \frac{s}{R^2}$.  In the limit $\La \rightarrow \infty$, $c_n,$ with $n \geq 1$ vanish. Therefore, let us focus on the divergent part ($c_{-1}$) and the finite part ($c_0$). In this regard, it is useful to note that the sum over $\ell$ can be expressed in terms of the third Jacobi theta function:
 \be \label{jacobi_theta}
 \sum_{\ell} e^{-s\ell^2} = \theta_3 \Big(0,q=e^{-s}\Big)
 \ee 
 In order to extract the divergent part of $I(\ep)$, we consider the limit $s \rightarrow 0$. Here, we will use the identity
 \be
\theta_3 \Big(0,e^{-s}\Big) =  \sqrt{\frac{\pi}{s}} \theta_3 \Big(0,e^{- \pi^2 / s}\Big)
 \ee 
 to get
 \be
 \sum_{\ell} e^{-s\ell^2}  = \sqrt{\frac{\pi}{s} } \sum_{n } e^{-\pi^2 n^2 /s} 
 \ee 
 It is clear that as $ s \rightarrow 0$, the leading contribution comes only from $n=0$. Using this fact in \eref{gap_int3} and noting that the divergent part comes from $\la_0 \rightarrow \infty$, we can replace $\tanh(\pi \la_0) \rightarrow 1$. We can now perform the $\la_0$-integral to find:
 \be
 \frac{\sqrt{\pi}}{2 R} \int_{\ep}^{\infty} \frac{ds} {s^{\frac32}}
 \ee
 where we have suppressed a factor of $2\pi$.
 It is now easy to perform the integral over $s$ and we find that $c_{-1}= \sqrt{\pi}$.
 \\
 \\
 To get the finite part of \eref{gap_int3} let us define:
 \be \label{finitepart}
 c_0 = \int_{0}^{\infty} ds\int  d \la_0 \la_0 \Bigg [\sum_{n=-\infty}^{\infty} \tanh(\pi \la_0) e^{-s(n^2 +\la_0^2)} -\int d \eta e^{-s(\la_0^2+\eta^2)} \Bigg]
 \ee 
 where we have subtracted the divergent part. We first perform the s-integral and using the identities
\begin{align}
 \int_{-\infty}^{\infty} d \eta \frac{1}{\eta^2 +\la_0^2} & = \frac{\pi}{\la_0} \\
 \sum_{n=-\infty}^{\infty} \frac{1}{n^2+\la_0^2} & = \frac{\pi coth(\pi \la_0)}{\la_0}  
\end{align}
we see that the finite part of \eref{finitepart} vanishes!
\\
In order to know the corrections to the trace-term \eref{gap_int3} slightly away from the conformal point $\bt= R$, we compute its derivative w.r.t $\bt$.
\be \label{bt_derivative}
\del_{\bt} I(\ep) \vert_{\bt=R} = \frac{1}{2\pi R^2} \int_{(\La R)^{-2}}^{\infty} ds \int d \la_0 \la_0 \tanh(\pi \la_0) e^{-s\la^2} \Bigg[-1 + 2 \sum_{\ell=1}^{\infty} (2 \ell^2 s-1)e^{-s\ell^2} \Bigg]
\ee 
We can recast the term inside square brackets as:
\be \label{sqr_bkt}
\sum_{\ell=-\infty}^{\infty} (2 \ell^2 s-1)e^{-s\ell^2} = -(2s \del_s +1) \theta_3\Big(0,q=e^{-s}\Big)= (2s e^{-s} \del_q -1) \theta_3\Big(0,q=e^{-s}\Big)
\ee
where we have used \eref{jacobi_theta}. Using the power series expansion of $\theta_3$, we find that as $s \rightarrow \infty$, $q \rightarrow 1$, and the derivative of $\theta_3$ is finite and the sum in \eref{sqr_bkt} becomes equal to $-1$. For $s=0$, $\theta_3$ diverges. To understand this better, let us use the modular transformation to write \eref{sqr_bkt} as 
\be
 -\lim_{s \rightarrow 0} (2s \del_s +1) \sqrt{\frac{\pi}{s}}  \theta_3\Big(0,q'=e^{-\frac{\pi^2}{s}}\Big) = -\lim_{s \rightarrow 0} 2 \sqrt{\frac{\pi}{s}} \Big[2q'\frac{\pi^2}{s}+O(q'^2)\Big]
\ee 
Thus, we see that the sum in \eref{sqr_bkt} is negative for all non-zero values of $s$ and converges to 0 as $s \rightarrow 0$. This, in particular, implies that \eref{bt_derivative} is negative.

Using these results in \eref{gap_int} the second gap equation becomes,
\be
\p^2 - \s + \frac{\sqrt{\pi}}{2\pi} \La + I_{finite} =0 
\ee 
where $I_{finite} = - c (\beta - R)$ with $c$ being a real, positive number. When $\phi = 0$, this implies that 
\be
	\s = \frac{\sqrt{\pi}}{2\pi} \La - c(\beta - R)
\ee

	\bibliographystyle{JHEP}
	\bibliography{renyion}
	
\end{document}